\documentstyle[prl,aps,epsf,epsfig,multicol]{revtex}

\begin{document}

\title{Testing a new Monte Carlo Algorithm for Protein Folding} 

\author{Ugo Bastolla$^1$, Helge Frauenkron$^1$, Erwin Gerstner$^{1, 2}$, 
      Peter Grassberger$^{1, 2}$, and Walter Nadler$^1$}

\address{ $^1$ HLRZ c/o Forschungszentrum J\"ulich, D-52425 J\"ulich, Germany\\
$^2$ Physics Department, University of Wuppertal, D-42097 Wuppertal, Germany}

\date{\today}

\maketitle

\begin{abstract}

  We demonstrate that the recently proposed pruned-enriched Rosenbluth 
  method PERM (P.~Grassberger, Phys. Rev. {\bf E}, in press (1997) ) leads to 
  extremely efficient algorithms for the folding of simple model proteins. 
  We test it on several models for lattice heteropolymers, and compare 
  to published Monte Carlo studies of the properties of particular sequences. 
  In {\it all} cases our method is faster than the previous ones, 
  and in several cases we find new minimal energy states. 
  In addition to producing more reliable candidates for ground states, 
  our method gives detailed information about the thermal spectrum and, thus,  
  allows to analyze static aspects of the folding behavior of arbitrary 
  sequences.
 
\end{abstract}

\pacs{87.15.By, 87.10.+e, 02.70.Lq}

\begin{multicols}{2}

\narrowtext

\section*{Introduction}
\label{sec:intro}

Protein folding \cite{Gierasch:King:90,Creighton:92,Merz:LeGrand:94,Brunak:95} 
is one of the most interesting and challenging  problems in polymer physics and 
mathematical
biology.  It is concerned with the problem of how a heteropolymer of a given 
sequence 
of amino acids folds into precisely that geometrical shape in which it 
performs its biological function as a molecular machine 
\cite{Drexler:86,Gross:95}. 
Currently, it is much simpler to find coding DNA --- and, thus,
also amino acid --- sequences than to elucidate the 3-$d$ structures of given 
proteins. 
Therefore, solving 
the protein folding problem would be a major break-through in understanding 
the biochemistry of the cell, and, furthermore, in designing artificial 
proteins. 

In this contribution we are concerned with the direct approach: 
given a sequence of amino acids, a molecular potential, 
and no other information, find the ground
state and the equilibrium state at physiological temperatures. 
Note that we are not concerned with the kinetics of 
folding, but only in the final outcome. Also, we will not address 
the problems of how to find good molecular potentials
\cite{Crippen:Maiorov:94,Kolinski:Skolnick:95,Mirny:Shakhnovich:96},
and what is the proper level of detail in describing proteins
\cite{Kolinski:Skolnick:95}. 
Instead, we will use simple coarse-grained models which have been 
proposed in the literature and have become standards in testing the efficiency 
of folding algorithms. 

A plethora of methods have been proposed to solve this problem, ranging 
from simple Metropolis Monte Carlo simulations at some nonzero temperature
\cite{Sali:Shakhnovich:Karlplus:94:JMB}
over multi-canonical simulation approaches
\cite{Hansmann:Okamoto:94-96}
to stochastic optimization schemes based, e.g., 
on simulated annealing \cite{Wilson:Cui:94}, 
and genetic algorithms \cite{Unger:Moult:93,LeGrand:Merz:94a}.
Alternative methods use 
heuristic principles \cite{Dill:Fiebig:Chan:93},
information from databases of known protein structures, 
\cite{Eisenhaber:Persson:Argos:95},
sometimes in combination with known physico-chemical properties
of small peptides.

The algorithms we apply here are variants of the pruned-enriched Rosenbluth 
method (PERM) \cite{alg}. This is a chain growth approach based on 
the Rosenbluth-Rosenbluth (RR) \cite{rr} method.
Preliminary results have been published before \cite{Frauenkron:etal:97}.
Here we will provide more details on the algorithm and on the
analyses that can be performed, and we will present more detailed results
on ground state and spectral properties, and on the folding behavior of
the sequences analyzed.

\section*{The Models}
\label{sec:models}

The models we study in this contribution are heteropolymers that live on 2- 
and 3-dimensional regular lattices. They are self-avoiding chains with attractive 
or repulsive interactions between neighboring non-bonded monomers.  

The majority of authors considered only two kinds of monomers. Although 
also different interpretations are possible for such a binary choice, e.g. 
in terms of positive 
and negative electric charges \cite{kantor-kardar:94}, the most 
important model of this class is the HP model \cite{Dill:85,Dill:89-91}. There, 
the two monomer types are assumed to be hydrophobic (H) and polar (P), with
energies $\epsilon_{HH}=-1,\; \epsilon_{HP}=\epsilon_{PP}=0$ for interaction 
between not covalently bound neighbors. Since this parameter set 
leads to highly degenerate  ground states, 
alternative parameters were proposed, e.g. 
$\boldmath{\epsilon}=(-3,-1,-3)$ \cite{socci1} and  
$\boldmath{\epsilon}=(-1,0,-1)$ \cite{otoole}. Note, however, 
that in these latter parameter sets, since they are
symmetric upon exchange of H and P, the intuitive distinction between
hydrophilic and polar monomers gets lost.

An interesting extension to the HP model can be obtained by allowing
the interactions to be anisotropic. This is done by introducing amphipatic (A)
monomers that have hydrophobic as well as polar sides 
\cite{Krausche:Nadler:97a}. 
Such a generalization is possible for all lattice types, but we confine 
ourselves to two dimensions (2d) here.
It can be shown that in this HAP model for a wide range of interaction 
parameters 
the inverse folding problem --- i.e. the determination of a sequence that has a 
particular
conformation as ground state ---
can be solved by construction \cite{Krausche:Nadler:97a}.
While 3-dimensional chiral amphipatic monomers can be considered as well as 
non-chiral (the sides are allowed to rotate in the latter),
in 2d only non-chiral monomers are possible.

In the other extremal case of models, 
all monomers of a sequence are considered to be different, and
interaction energies are drawn randomly from a continuous distribution
\cite{Sali:Shakhnovich:Karplus:94:Nature,Klimov:Thirumalai:96}.
These models correspond, effectively, 
to assuming an infinite number of monomer types.

\section*{The Sequences}
\label{sec:sequences}

For the above models various sequences were analyzed in the literature,
and in \cite{Frauenkron:etal:97}
we took these analyses as a test for our algorithm.
Here we will take a closer look at the properties of some
of the sequences that were considered there.

\end{multicols}
\widetext

\vglue-.2cm
\begin{table}
\caption{Newly found lowest energy states for binary sequences with 
interactions $\epsilon = (\epsilon_{HH},\epsilon_{HP},\epsilon_{PP})$. 
Configurations are encoded as sequences of {\it r}(ight),{\it l}(eft),
{\it u}(p), {\it d}(own), {\it f}(orward), and {\it b}(ackward).}
\label{seq.struct.table}

\begin{tabular}{cccccc} 
  $N$ & $d$ & $\epsilon$ &    sequence   & old $E_{\rm min}$ & Ref. \\ 
      &     &            & configuration & $E$      &       \\ \hline

 100  & 2 & $(-1,0,0)$ &
$P_6HPH_2P_5H_3PH_5PH_2P_2(P_2H_2)_2PH_5PH_{10}PH_2PH_7P_{11}H_7P_2HPH_3P_6HPH_2
$ & $-44$ & \cite{pekney} \\
 & & & 
$r_6ur_2u_3rd_5luldl_2drd_2ru_2r_3(rulu)_2urdrd_2ru_3lur_3dld_2rur_5d_3l_5uldl_2
d_3ru_2r_3d_3l_2urul$ 
& $-47$ & \\
 & & &
$rdldldrd_2r_2d_3l_2drdldr_2dl_2dl_2(urul)_2urur_2ul_2u_2l_2drd_2lul_2uru_2r_2u_
4rul_3drd_3l_2d_2-$ & & \\
 & & & $ ldlu_2ru_2lu_3rd_2rdr$ & $-47$ & \\
 & & &
$u_3r_3ur_5dl_4drd_2r_2ulur_4dr_2dld_2lu_2luld_2rdl_2ul_3dr_2dr_2drurdr_3d_2luld
_2lu_2l_2drd_2lulu_3l_2ul_2u_3ru$
& $-47$ & \\
 & & &
$rdr_3dldl_3drdrur_2ur_2ulur_2urd_2(ldrd)_2l_2u_2ldl_2dr_2dr_3dl_2dlulul_2ul_2d_
3ldr_3u_2rdrdldl-$ & & \\
 & & & $ dr_2urur_4u_3rul$ & $-47$ & \\
 & & &
$rd_3rdldl_2uld_3ld_2rur_3dl_2dr_6ul_3u_2l_3ur_3ur_2dld_2rur_2ulu_2l_2u_3lu_3r_2
d_5r_2d_2rdru_2lu_2r_2u_2ldl_2dl$
& $-47$ & \\
 & & &
$r_3u_3ru_2ru_3l_2ur_2ul_2urul_4dr_2dldrdld_2rur_2dld_2luld_2rdl_3uru_2ldld_3l_3
ururu_2rur_2ulu-$ & & \\
 & & & $ ldldl_2u_3ld_4r$ & $-47$ & \\\hline

 100  & 2 & $(-1,0,0)$ &
$P_3H_2P_2H_4P_2H_3(PH_2)_3H_2P_8H_6P_2H_6P_9HPH_2PH_{11}P_2H_3PH_2PHP_2HPH_3P_6
H_3$ & $-46$ & \cite{pekney} \\
 & & &
$ru_2ldlu_2ld_2lu_2lurulur_2d_2ru_4r_2dld_4ru_3rdrdld_6rdru_3rul_2u_2rdr_2ululu_2
(rd)_3rur_2dldld_4lulu_2ru$
& $-49$ & \\
 & & &
$u_3rdru_2rd_2ru_2r_2u_2ldluldlu_5ld_6l_2d_2lu_3r_2u_6l_2d_3ldr_2dl_2dldlu_2ru_2l
dl_2drdl_2d_2rurdrd_3ru_3ru$
& $-49$ & \\
 & & &
$ul_2drdl_2u_3ld_4ldrdl_2u_2l_2d_3l_2uru_3r_2u_3rd_3ru_4rul_5dldr_2d_2luldldrdldl
u_3lul_2ulur_2dr_2u_3rd_4l$
& $-49$ & \\ \hline


  60  & 2 & $(-1,0,0)$ &
  $ P_2H_3PH_8P_3H_{10}PHP_3H_{12}P_4H_6PH_2PHP $ & $-34$ & 
\cite{Unger:Moult:93} \\
 & & & $r_5d_2lul_3dld_2(ru)_2rd_2ldldrdr_2uluru_2rd_2rdldr_2u_3lu_3rd_2rur$ 
&$-36$ &  \\ \hline

  80  & 3 & $(-1,0,-1)$ & 
 ${PH_2P_3(H_3P_2H_3P_3H_2P_3)_3H_4P_4(H_3P_2H_3P_3H_2P_3)H_2}$ & $-94$ & 
\cite{otoole,deutsch} \\
 & & & 
$lbruflbl_2br_2drur_2dldl_3ulfrdr_3urfldl_3ulurur_3drblul_3br_3bl_3dldrdr_3urul_
2dlu 
$ & $-98$ &  \\ 
\end{tabular}
\end{table}

\begin{multicols}{2}
\narrowtext

\subsection*{2d HP model}
\label{sec:sequences:2dHP}

Two-dimensional HP chains were used in several papers as test cases for 
folding algorithms. We shall discuss the following 
ones:

(a) Several chains of length 20 to 64 were studied in \cite{Unger:Moult:93} 
by means of a genetic algorithm. These authors quote supposedly exact 
ground state energies, and lowest energies obtained by simulations. While 
these coincide for the shorter chains ($N\leq 50$), the authors were unable 
to fold the longer chains with $N=60$ and $64$.

\begin{figure}
\begin{center}
\epsfig{figure=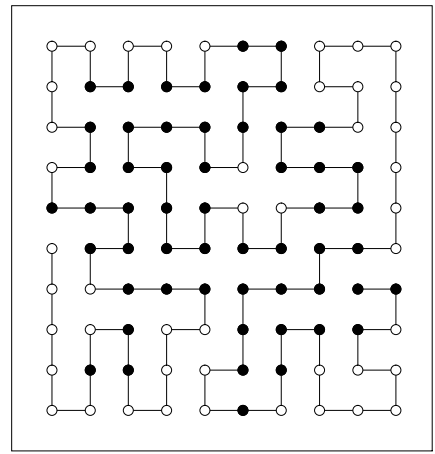, width=2.2truein}
\end{center}
\caption{ Putative compact native structure of sequence 1 from Table~I ($E=-45$)
according to \protect\cite{pekney};
(filled circle) H monomers, (open circle) P monomers.
}
\label{rama1.struct}
\end{figure}

\begin{figure}
\begin{center}
\epsfig{figure=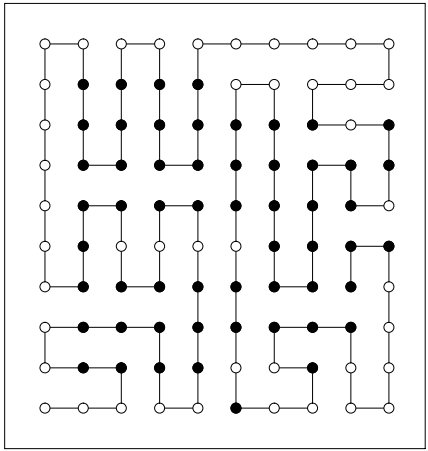, width=2.2truein}
\end{center}
\caption{ Putative compact native structure of sequence 2 from Table~I ($E=-46$) 
according to \protect\cite{pekney}.
}
\label{rama2.struct}
\end{figure}

(b) Two chains with $N=100$ were studied in 
\cite{pekney}. The authors claimed that their native configurations 
were compact, fitting exactly into a $10\times 10$ square, and had energies 
$-44$ and $-46$, see Table~I for the sequences and Fig.~\ref{rama1.struct} 
and \ref{rama2.struct} for the respective proposed ground state structures.  
These conformations were found by a specially 
designed MC algorithm which should be particularly efficient for compact 
configurations. 

\subsection*{2d HAP model}
\label{sec:sequences:2dHAP}

\begin{figure}
\begin{center}
\epsfig{figure=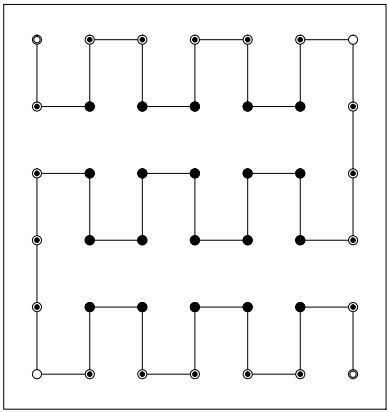, width=2.truein}
\end{center}
\caption{ Ground state structure of the HAP sequence;
(filled circle) H monomers, (structured circle) A monomers,
(open circle) P monomers.
}
\label{HAP.42.struct}
\end{figure}

To analyze the performance of PERM on the HAP model we 
used $\boldmath{\epsilon}=(-4,-1,2)$ as energy parameters, for which 
set the inverse folding problem is solvable \cite{Krausche:Nadler:97a}.
We choose a 3-helix structure with \hbox{$N=42$}, see 
Fig.~\ref{HAP.42.struct},
which is the ground state of the sequence $pA(HAAH)_3PA_3H_{10}A_3P(HAAH)_3Ap$,
where $A$ denotes an amphipatic intra-chain group with one hydrophobic 
and one polar side, while $p$ denotes an amphipatic end group with one 
hydrophobic and two polar sides. 

\subsection*{3d HP model}
\label{sec:sequences:3dHP}

Ten sequences of length $N=48$ were given in \cite{yue-shak}. 
Each of these sequences was designed by 
minimizing the energy of a particular target conformation  
in sequence space under the constraint of constant composition
\cite{Shakhnovich:93-94}.
The authors tried to find the
lowest energy states with two different methods,
one being an heuristic stochastic approach \cite{Dill:Fiebig:Chan:93}, 
the other based on exact enumeration of
low energy states \cite{Yue:Dill:95}. 
With the first method they failed in all but one case 
to reach the lowest energy. With the second method in all but one cases 
they obtained conformations with energies that were even lower than 
the putative ground states the sequences were designed for, while
for one case the ground state energy was confirmed.
Precise CPU times were not quoted.

\subsection*{3d modified HP model}
\label{sec:sequences:3dmodHP}

\begin{figure}
\epsfig{figure=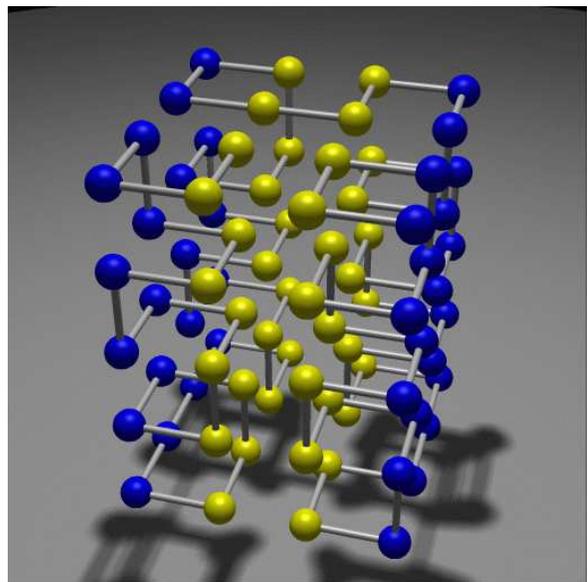, width=3truein}
\caption{ Putative native state of the ``four helix bundle" 
sequence, see Table~I, as proposed by O'Toole {\it et al.}. It has $E=-94$, fits 
into 
a rectangular box, and consists of three homogeneous layers. Structurally, 
it can be interpreted as four helix bundles.} 
\label{3d.struct.old}
\end{figure}

A most interesting case is a 2-species 80-mer with interactions 
$(-1,0,-1)$ studied first in \cite{otoole}. These particular 
interactions were chosen instead of the HP choice $(-1,0,0)$ because it 
was hoped that this would lead to compact configurations. Indeed, the 
sequence was specially designed to form a ``four helix bundle" 
which fits perfectly into a $4\times4\times5$ box, see Fig.~\ref{3d.struct.old}. 
Its energy in this putative native state is $-94$.
Although the authors of \cite{otoole} used highly optimized codes, they 
were not able to recover this state by MC. Instead, they reached only 
$E=-91$. Supposedly, a different state with $E=-94$ was found in 
\cite{pekney}, but figure 10 of this paper, which is claimed to 
show this configuration, has a much higher value of $E$. Configurations 
with $E=-94$ but slightly different from that in \cite{otoole} and 
with $E=-95$ were found in \cite{deutsch} by means of an algorithm similar
to that in \cite{pekney}. For each of these low energy states the 
author needed about one week of CPU time 
on a Pentium.

\subsection*{3d $\infty$ monomer types}

Sequences with $N=27$ and with continuous interactions were studied 
in \cite{Klimov:Thirumalai:96}. Interaction strengths were sampled from 
Gaussians with fixed non-zero mean and fixed variance. These $N(N-1)/2$ 
numbers were first attributed randomly to the monomer pairs,
then they were randomly permuted, using a Metropolis accept/reject 
strategy with a suitable cost function, to obtain good folders. 
Such ``breeding" strategies to obtain good folders 
were also developed and employed by other authors for various models
\cite{Shakhnovich:93-94,Ebeling:Nadler:95-97,deutsch-kurosky:96}, 
and seem necessary to eliminate sequences which fold too slowly and/or 
unreliably. It is believed that also during biological evolution optimization 
processes took place with similar effects, so that actual proteins are 
better folders than random amino sequences.

\section*{The Algorithm}
\label{sec:algorithm}

The algorithms we apply here are variants of the pruned-enriched Rosenbluth 
method (PERM) \cite{alg}, a chain growth algorithm based on 
the Rosenbluth-Rosenbluth (RR) \cite{rr} method. There, monomers are 
placed sequentially at vacant sites either with uniform probability, or 
with some non-uniform probability distribution. In either case it leads 
to weighted samples where each configuration carries a weight $W$. For 
long chains or low temperatures, the spread in weights can become very 
wide which then leads to serious problems \cite{kremer}. But
since the weights accumulate as the chains grow, one can interfere 
during the growth process by `pruning' configurations with low weights 
and replacing them by copies of high-weight configurations. This is in 
principle similar to population based methods in polymer simulations 
\cite{garel,velicson} and in quantum Monte Carlo (MC) \cite{umrigar}, 
but the implementation is different. Pruning is done stochastically: 
if the weight of a configuration has decreased below a threshold $W^<$, 
it is eliminated with probability 1/2, while it is kept and its weight is 
doubled in the other half of cases. Copying (`enrichment' \cite{enrich}) 
is done independently of this. If $W$ increases above another threshold $W^>$, 
the configuration is replaced by $n$ copies, each with weight $W/n$. 
Technically, this is done by putting onto a stack all information needed
about configurations which still have to be copied. This is most easily 
implemented by recursive function calls. Thereby one avoids the need for 
keeping large populations of configurations \cite{garel,velicson,umrigar}. 
PERM has proven
extremely efficient for studies of lattice homopolymers near the $\theta$ 
point \cite{alg}. It has also been successfully applied to phase equilibria 
\cite{multic}, to the ordering transition in semi-stiff polymers 
\cite{stiff}, and to spiraling transitions of polymers with interactions 
depending on relative orientation of monomers \cite{spiral}. We refer to 
these papers for more detailed descriptions of the basic algorithm. 

The main freedom when applying PERM consists in the a priori choice of 
the sites where to place the next monomer, in the thresholds $W^<$ 
and $W^>$ for pruning and copying, and in the 
number of copies made each time. All these features do not affect the
formal correctness of the algorithm, but they can greatly influence its 
efficiency. They may depend arbitrarily on chain lengths and on 
local configurations, and they can be changed freely at any time during 
the simulation. Thus the algorithm can `learn' during the simulation.

In order to apply PERM to heteropolymers at very low temperatures, 
the strategies proposed in \cite{alg} are modified as follows.

(1) For homopolymers near the theta-point 
it had been found that the best choice for the placement of 
monomers was not according to their Boltzmann weights, but uniformly on 
all allowed sites 
\cite{alg,multic}. This might be surprising since the Boltzmann factor has 
then to be included into the weight of the configuration, which might 
lead to large fluctuations. Obviously, this effect is 
counterbalanced by the fact that larger Boltzmann factors correspond 
to higher densities and thus to smaller Rosenbluth factors \cite{kremer}. 

For a heteropolymer this has to be modified, as there is no longer a 
unique relationship between density and Boltzmann factor. In a strategy 
of `anticipated importance sampling' we should preferentially 
place monomers on sites with mostly attractive neighbors. Assume that we 
have two kinds of monomers, and we want to place a type-$A$ monomer. If 
an allowed site has $m_B$ neighbors of type $B$ ($B=H,P$), we select 
this site with a probability $\propto 1+a_{AH}m_H+a_{AP}m_P$. Here, $a_{AB}$ 
are constants with $a_{AB}>0$ for $\epsilon_{AB}<0$ and vice versa. 

(2) Most naturally, $W^>$ and $W^<$ are chosen proportional to the 
estimated partition sum $Z_n$ 
\cite{alg}. This becomes inefficient at very low $T$ since $Z_n$ will 
be underestimated as long as no low-energy state is found. When this 
finally happens, $W^>$ is too small. Thus too many copies are made which 
are all correlated but cost much CPU time. 

This problem can be avoided by increasing $W^>$ and $W^<$ during particularly 
successful `tours' (a tour is the set of configurations derived by copying 
from a single start \cite{alg}). But then also the average number of long 
chains is decreased in comparison with short ones. To reduce this effect 
and to create a bias towards a sample which is flat in chain length, 
we multiply by some power of $M_n/M_1$, where $M_n$ is the number of 
generated chains of length $n$. With ${\cal N}(n)$ denoting the number 
of chains generated during the current tour we used therefore
$$
   W^< = C\;Z_n\; [(1+{\cal N}(n)/M)(M_n+M)/(M_1+M)]^2 , 
$$
and $W^> = rW^<$. Here, $C$ is a constant of order unity, $r\approx 10$, 
and $M$ is a constant of order $10^4 - 10^5$.

(3) Creating only one new copy at each enrichment event (as done in 
\cite{alg}), cannot prevent the weights from exploding at very low $T$. 
Thus we have to make several copies if the weight is large and surpasses 
$W^>$ substantially. A good choice for the number of new copies created when 
$W>W^>$ is int$[\sqrt{W/W^>}]$.

(4) Two special tricks were employed for `compact' configurations of the 2-$d$ 
HP model filling a square. First of all, since we know in this case where 
the boundary should be, we added a bias for polar monomers to actually 
be on that boundary, by adding an additional energy of -1 per boundary site. 
Note that this bias has to be corrected in the weights, thus the final distributions are 
unaffected by it and unbiased. Secondly, in two dimensions we can immediately 
delete chains which cut the free domain into two disjoint parts, since they 
never can grow to full length. In the present simulations, we checked for 
this by looking ahead one time step. In spite of the additional work this 
was very efficient, since it reduced considerably the time spent on dead-end 
configurations.

(5) In some cases we did not start to grow the chain from one end but 
from a point in the middle. We grew first one half, and then the other. 
Results were averaged over all possible starting points.
The idea behind this is that real proteins have folding nuclei
\cite{Matheson:Scheraga:78}, and it should be most efficient to
start from such a  nucleus. In some cases this trick was very successful and 
speeded up the ground state search substantially, in others not. 
We take this observation as an indication that in various sequences 
the end groups already provide effective nucleation sites. This is e.g. 
the case for the 80-mer with modified HP interactions of \cite{otoole}.
We also tried to grow the chain on both sides simultaneously. However
it turned out that this is not effective computationally 
\cite{Grassberger:unpublished}.

(6) In the case of the HAP model 
it turned out that, while the ground state configuration of the chain geometry
could be reached easily, the ground state configurations of the side groups
(i.e., of the H/P bond attributions of amphiphilic monomers 
\cite{Krausche:Nadler:97a}) could note be reached effectively using PERM alone. 
Therefore, after building the chain conformation to its full length
we let the side groups of the amphipatic monomers rotate thermally 
using a Metropolis algorithm. 
This approach utilizes the short relaxation time of side group fluctuations
within the subphase of a fixed chain conformation and leads 
to the desired ground states
\cite{enumerationNote}.

(7) For an effective sampling of low-lying states the choice of 
simulation temperature $T$
appears to be of importance. If it is too large, low-lying states will 
have a low statistical weight and will not be sampled reliably. On the 
other hand, if $T$ is too low, the algorithm becomes too greedy: configurations 
which look good at first sight but lead to dead ends are sampled too often, 
while low energy configurations, whose qualities become apparent only at 
late stages of the chain assembly, are sampled rarely. Of course they then 
get huge weights (since the algorithm is correct after all), but statistical 
fluctuations become huge as well. This is in complete analogy to the 
slow relaxation hampering more traditional (Metropolis type) simulations 
at low $T$  -- note, however, that ``relaxation" in the proper sense does not 
exist in the present algorithm.

In the cases we considered it turned out to be most effective to choose
a temperature that is below the collapse transition temperature
(note, however, that this transition is smeared out, see the results below)
but somewhat above the temperature corresponding to the structural transition 
which leads to the native state.
This observation corresponds qualitatively to the considerations of
\cite{Finkelstein:Gutin:Badretdinov:95},
although a quantitative comparison appears not to be possible.

(8) For 2-dimensional HP chains, we performed also some runs where we restricted the 
search for native configurations further, by disallowing non-bonded HP 
neighbor pairs. The idea behind this was that such a pair costs energy, 
and is thus less likely to appear in a native state. But this is only 
a weak heuristic argument. Forbidding such pairs certainly gives wrong 
thermal averages, and it might prevent the native state to be found, if 
it happens to contain such a pair. But in two cases this restriction did work, and 
gave states with lower energies than those we could reach without this 
trick.

\section*{Results}
\label{sec:results}

Let us now discuss our results.
All CPU times quote below refer to SPARC Ultra machines with 167 MHz.

\subsection*{2d HP model}
\label{sec:results:2dHP}

\begin{figure}
\begin{center}
\epsfig{figure=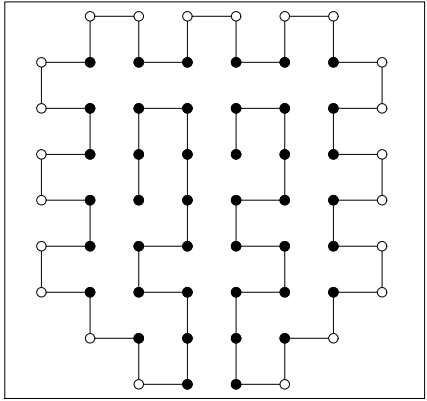,angle=270, width=2.2truein}
\end{center}
\caption{ One of the ground states of the $N=64$ sequence of 
\protect\cite{Unger:Moult:93}. The other ground states differ in the configurations of 
the chain ends filling the interior of the structure, but have the same boundary 
and overall shape.}
\label{unger64}
\end{figure}

(a) For all chains of \cite{Unger:Moult:93} we easily reached the ground state, 
except for the longest chain ($N=64$). For this chain the ground state is 
very regular, with all polar monomers on the outside, and with no non-bonded 
HP neighbors. Its energy is -42 (see fig.\ref{unger64}). The authors of 
\cite{Unger:Moult:93} were unable to recover by simulations this state 
and any other state with $E<-37$. Although we could not reach $E=-42$ either 
without special tricks, we obtained at least $E=-40$ after ca. 4h CPU time. 
Forbidding non-bonded HP neighbor pairs, we found the native state easily, 
but this cannot be really counted as a success since we knew in advance that 
such pairs do not appear in the native state. The difficulty posed by this 
sequence for PERM is obvious from fig.\ref{unger64}: there is no folding center 
in this chain. No matter where one starts the assembly, one first has to 
construct a large part of the boundary before the interior behind this 
boundary is filled up. This intermediate state has a very low Boltzmann 
weight and acts thus as a bottleneck for PERM. 

Note that the configuration of Fig.\ref{unger64} cannot be obtained by popular 
local moves including, e.g., bond exchange or crankshaft, or by reptation. But it 
can be reached if pivot moves are included. This illustrates that folding 
properties can depend strongly on the chosen kinetics.

In contrast to the sequence with $N=64$, we had no problem with the shorter
sequences of \cite{Unger:Moult:93}. In particular, for the $N=60$ sequence 
we found a configuration with $E=-36$ (see Table I), although the authors 
had quoted $E=-34$ as supposedly exact ground state energy. 

\begin{figure}
\begin{center}
\epsfig{figure=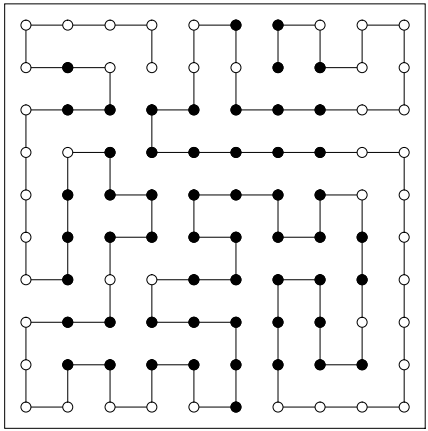, width=2.2truein}
\end{center}
\caption{ One of the compact structures of sequence 1 with energy 
($E=-46$) lower than the ``native" state proposed by Ramakrishnan {\it et al.}.  }
\label{2d.seq1.struct.compact}
\end{figure}

\begin{figure}
\begin{center}
\epsfig{figure=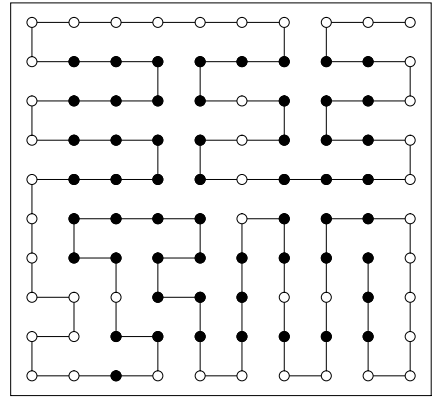, width=2.2truein}
\end{center}
\caption{ One of the compact structures of sequence 2 with lower energy 
($E=-47$).
}
\label{2d.seq2.struct.compact}
\end{figure}

\begin{figure}
\begin{center}
\epsfig{figure=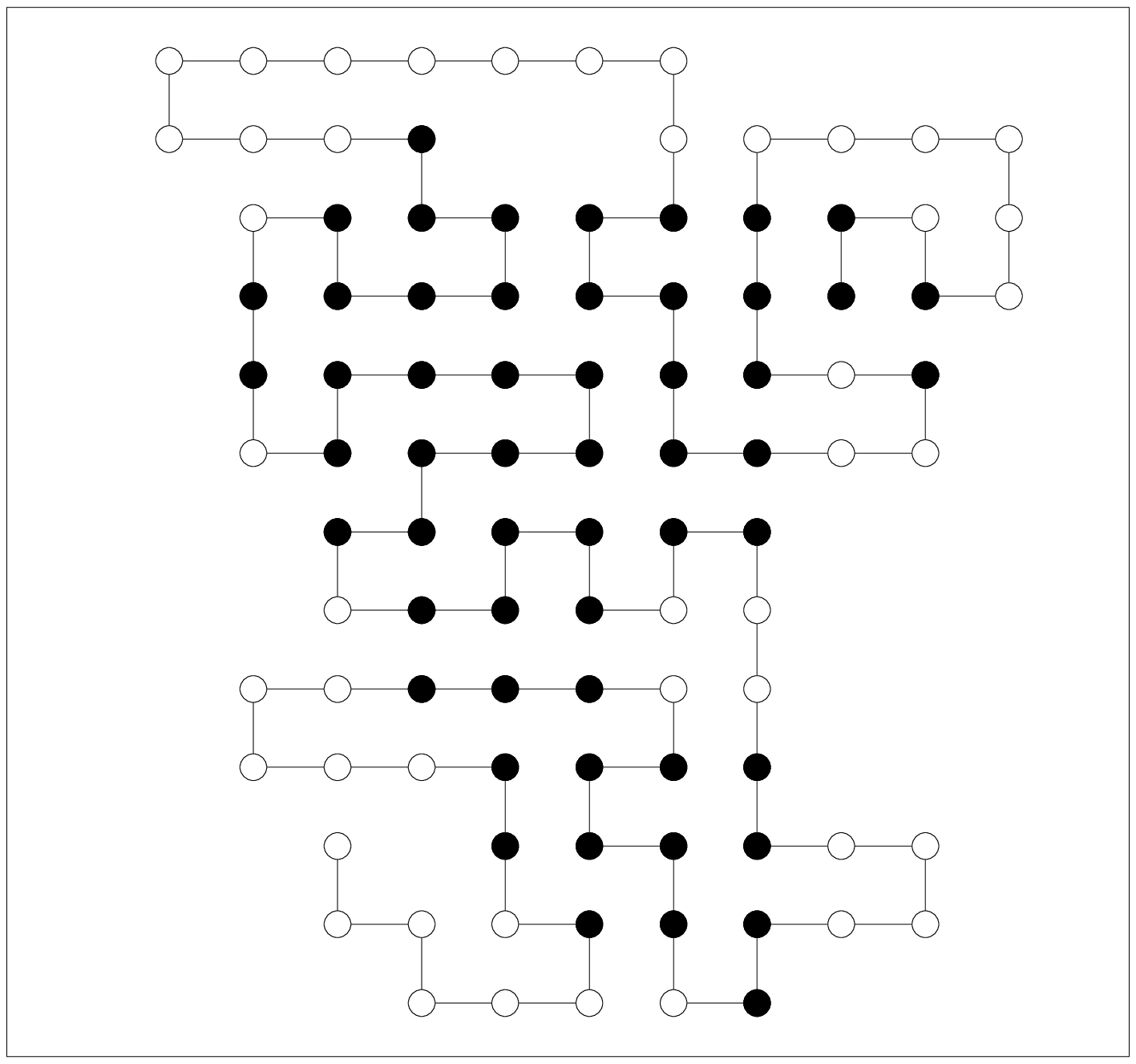, width=2.2truein}
\end{center}
\caption{ One of the (non-compact) lowest energy sequences for sequence 1 
($E=-47$).
}
\label{2d.seq1.struct.new}
\end{figure}

\begin{figure}
\begin{center}
\epsfig{figure=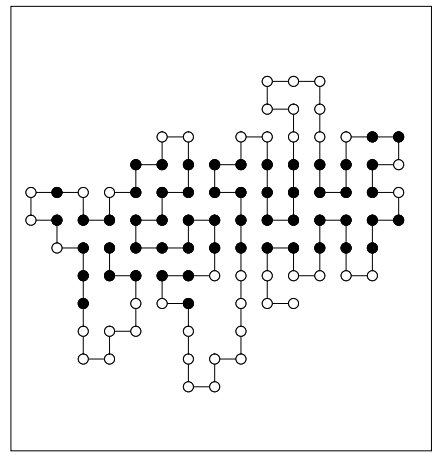, width=2.2truein}
\end{center}
\caption{ One of the (non-compact) lowest energy sequences for sequence 2 
($E=-49$). }
\label{2d.seq2.struct.new}
\end{figure}

(b) For the two HP chains of \cite{pekney} with $N=100$, see Table I, 
we found several compact states (within ca. 40 hours of CPU time) 
that had energies lower than those of the compact putative ground 
states proposed in \cite{pekney}.
Figures \ref{2d.seq1.struct.compact} and \ref{2d.seq2.struct.compact} 
show representative compact structures with $E=-46$
for sequence 1 and $E=-47$ for sequence 2.
Moreover, we found (again within 1-2 days of CPU time) several 
non-compact configurations with energies 
even lower: $E=-47$ and $E=-48$ for sequence 1 and 2, respectively. 
Forbidding non-bonded HP pairs, we obtained even $E=-49$ for 
sequence 2.
Figures \ref{2d.seq1.struct.new} and \ref{2d.seq2.struct.new} show 
representative non-compact structures with these energies; a
non-exhaustive collection of these is listed in

\noindent
Table~I. These results reflect the well-known property 
that HP sequences 
(and those of other models) usually have ground states
that are not maximally compact, see, e.g. \cite{yue-shak}, although 
there is a persistent prejudice to the contrary
\cite{otoole,pekney,Shakhnovich:Gutin:90-93}.

\subsection*{2d HAP model}
\label{sec:results:2dHAP}

The ground state of the HAP triple helix was found within
several minutes of CPU time
using the PERM-Metropolis hybrid algorithm. It was not found to be degenerate.

In order to obtain information about the folding transition, energy and 
contact matrix (see below) histograms were determined at $T=1.25,2$, and 4. Free 
energy differences, necessary for combining the histograms \cite{mchisto},
were determined using Bennet's acceptance ratio method \cite{Bennet:76}.

\begin{figure}
\begin{center}
\epsfig{figure=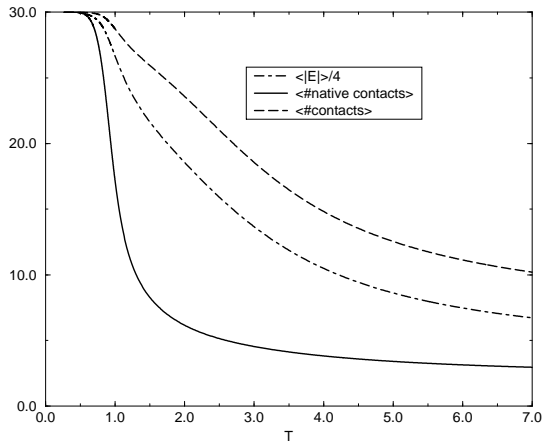,angle=270, 
width=2.8truein}
\end{center}
\caption{ 
Number of native contacts, total number of contacts, and energy $E$ $vs$ 
temperature $T$
for the HAP sequence of Fig. \protect\ref{HAP.42.struct}.
}
\label{HAP.42.op.vs.T}
\end{figure}

In Fig.~\ref{HAP.42.op.vs.T} the thermal behavior of the mean number of 
contacts, number of native contacts, and of the mean energy is shown.
While the structural transition to the ground state phase is best 
monitored by the number of native contacts and takes place around $T=1$, 
the compactification of the polymer chain is most clearly seen in the 
mean number of all contacts and in the radius of gyration (not shown here).
It takes place already at much higher temperatures and is smeared out over 
a wide temperature range. Note that the number of all contacts follows 
closely the behavior of the energy.

\begin{figure}
\begin{center}
\epsfig{figure=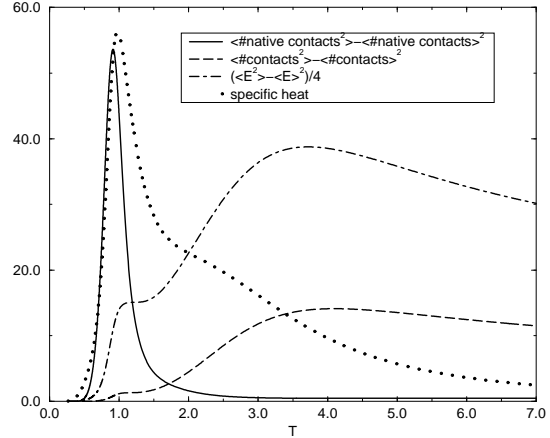, 
angle=270,width=2.8truein}
\end{center}
\caption{ 
Fluctuations of the number of native contacts, total number of contacts,
and energy $E$ $vs$ temperature $T$ for the HAP sequence of Fig. 
\protect\ref{HAP.42.struct};
Since energy fluctuations and specific heat emphasize the polymer collapse
and structural transition differently, we have included the specific heat, too.
}
\label{HAP.42.fluc.vs.T}
\end{figure}

These two transitions are seen more clearly when the fluctuations of the 
above order parameters are considered, see Fig.~\ref{HAP.42.fluc.vs.T}.
Energy and contact number fluctuations exhibit a broad maximum around 
$T=3.5$, while the structural transition is indicated by a narrow peak of
the fluctuations of the number of native contacts and of 
the specific heat near $T=1.0$. Note that specific heat 
and energy fluctuations emphasize 
the two transitions differently due to the factor $T^2$ by which they differ.

\subsection*{3d HP model}
\label{sec:results:3dHP}

\begin{figure}
\begin{center}
\epsfig{figure=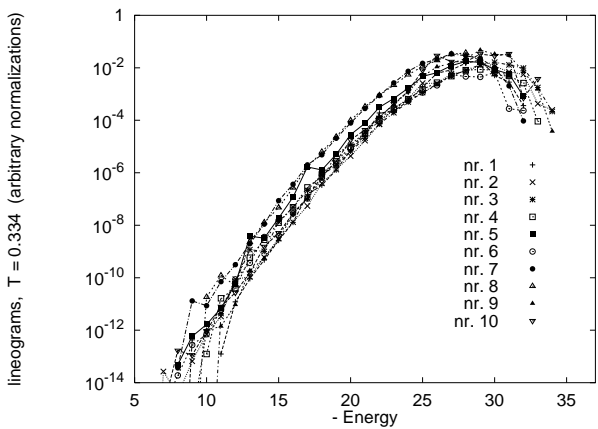, angle=270, width=8.3cm}
\end{center}
\caption{Energy spectrum of the ten sequences given in \protect\cite{yue-shak}. 
More precisely, to emphasize the low-energy part of the spectrum, we show the 
histograms obtained from the spectra by multiplying 
with \protect$e^{E/T}, \;T=0.334$. Note that there are no 
energy gaps in any of these spectra.
}
\label{NoGaps}
\end{figure}

With PERM we succeeded to reach ground states of the
ten sequences of length $N=48$ given in \cite{yue-shak}
in {\it all} cases, in CPU times between a few seconds and 5 hours, see 
Table II. In these simulations we used a rather simple version of PERM, 
where we started assembly always 
from the same end of the chain. We 
found that the sequences most difficult to fold were also those which 
had resisted previous Monte Carlo attempts \cite{yue-shak}.
In those cases where a ground state was hit more than once, we verified 
also that the ground states were highly degenerate. In no case there 
were gaps between ground and first excited states, see Fig.~\ref{NoGaps}.
Therefore, none of these sequences is a good folder,
though they were designed specifically for this purpose.

\subsection*{3d modified HP model}
\label{sec:results:3moddHP}

\begin{figure}
\epsfig{figure=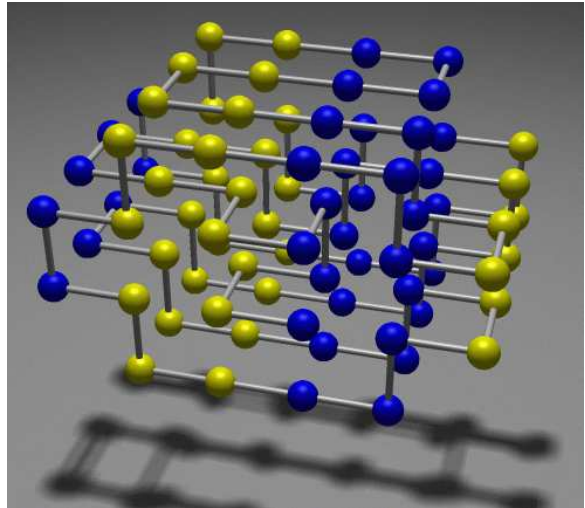, width=3truein}
\caption{ Conformation of the ``four helix bundle"
sequence with $E=-98$. We propose that this is the actual ground 
state. Its shape is highly symmetric although it does not fit into a 
rectangular box. It is not degenerate except for a flipping of the central 
front $2\times2\times2$ box.}
\label{3d.struct.new}
\end{figure}

For the two-species 80-mer with interactions 
$(-1,0,-1)$, even without much tuning our algorithm gave $E=-94$ after a few 
hours, but it did not stop there. After a number of rather disordered 
configurations with successively lower energies, the final candidate 
for the native state has $E=-98$. It again has a highly symmetric 
shape, although it does not fit into a $4\times4\times5$ box, see 
Fig.~\ref{3d.struct.new}. 
It has twofold degeneracy (the central $2\times2\times2$ 
box in the front of Fig.~\ref{3d.struct.new} can be flipped), and both 
configurations were actually found in the simulations. Optimal
parameters for the ground state search in this model are 
$\beta=1/kT\approx 2.0$, $a_{PP} = a_{HH} \approx 2$, 

\begin{minipage}{8.cm}
\begin{table}
\caption{ PERM performance for the binary sequences from \protect\cite{yue-shak}.
} \label{table2}
\begin{tabular}{ccccr}
 sequence & 
 $-E_{\rm min}$\tablenote{Ground state energies \protect\cite{yue-shak}.} &
 $-E_{\rm MC}$\tablenote{Previously reached energies with Monte Carlo methods 
      \protect\cite{yue-shak}.} & 
 $n_{\rm success}$\tablenote{Number of independent tours in which a ground state was hit.} & CPU time   \\
   nr.    &                &                  &               &(min)  \\ \hline
    1     &    32    &    31    &   101     &     6.9  \\
    2     &    34    &    32    &    16     &    40.5  \\
    3     &    34    &    31    &     5     &   100.2  \\
    4     &    33    &    30    &     5     &   284.0  \\
    5     &    32    &    30    &    19     &    74.7  \\
    6     &    32    &    30    &    24     &    59.2  \\
    7     &    32    &    31    &    16     &   144.7  \\
    8     &    31    &    31    &    11     &    26.6  \\
    9     &    34    &    31    &     1     &  1420.0  \\
   10     &    33    &    33    &    16     &    18.3  \\
\end{tabular}
\end{table}
\end{minipage}

\noindent 
and $a_{HP}\approx -0.13$. With these, 
average times for finding $E=-94$ and $E=-98$ in new tours are 
ca. 20 min and 80 hours, respectively.

A surprising result is that the monomers are arranged in four homogeneous 
layers in Fig.~\ref{3d.struct.new}, while they had formed only three 
layers in the putative ground state of Fig.~\ref{3d.struct.old}. 
Since the interaction should favor the segregation 
of different type monomers, one might have guessed that a configuration 
with a smaller number of layers should be favored. We see that this 
is outweighed by the fact that both monomer types can form large double 
layers in the new configuration. Again, our new ground state is 
not `compact' in the sense of minimizing the surface, and hence 
it also disagrees with the wide spread prejudice 
that native states are compact.

In terms of secondary structure, the new ground state
is fundamentally different from the putative ground state 

\begin{figure}
\begin{center}
\epsfig{figure=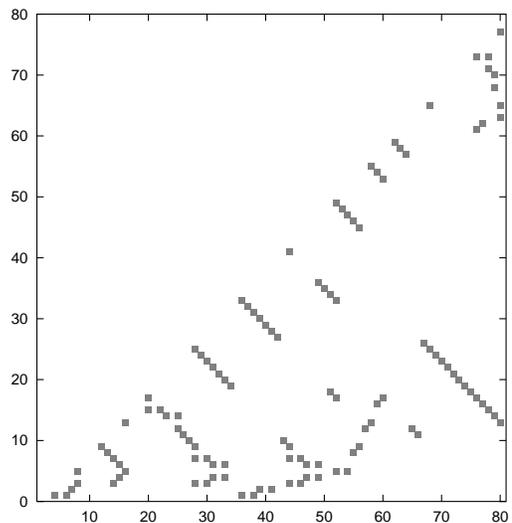, width=2.85truein}
\end{center}
\caption{ Contact matrix of the structure in Fig.~\protect\ref{3d.struct.new};
a black point at $(i,j)$ indicates that there is a contact between monomer $i$
and monomer $j$; grey points indicate contacts in only one of the two 
native states, corresponding to the twofold degeneracy of the central 
$2\times2\times2$ box.
Note that the lines orthogonal to the main diagonal correspond to 
anti-parallel $\beta$ sheet secondary structure elements,
see e.g. \protect\cite{Chan:Dill:89-90}.
}
\label{3d.contact.new}
\end{figure}

\begin{figure}
\begin{center}
\epsfig{figure=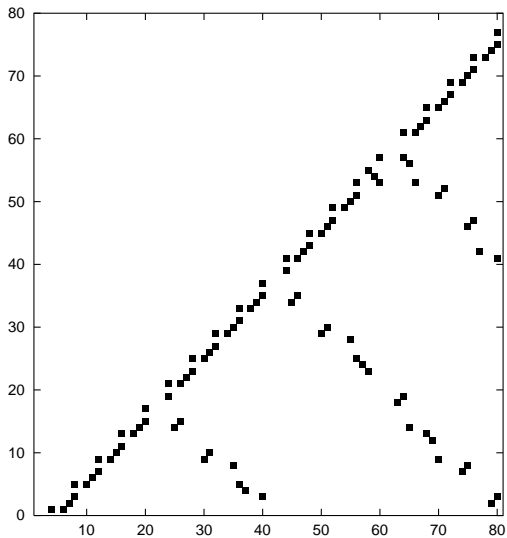, width=2.85truein}
\end{center}
\caption{ For comparison, the contact matrix of the 
putative ground state of Ref.~\protect\cite{pekney} in 
Fig.~\protect\ref{3d.struct.old};
note that point triples close to the diagonal parallel as well as orthogonal to 
it are signatures of 3d helical secondary structure elements,
see e.g. \protect\cite{Chan:Dill:89-90}; the other points 
denote tertiary contacts between helices.
}
\label{3d.contact.old}
\end{figure}

\begin{figure}
\epsfig{figure=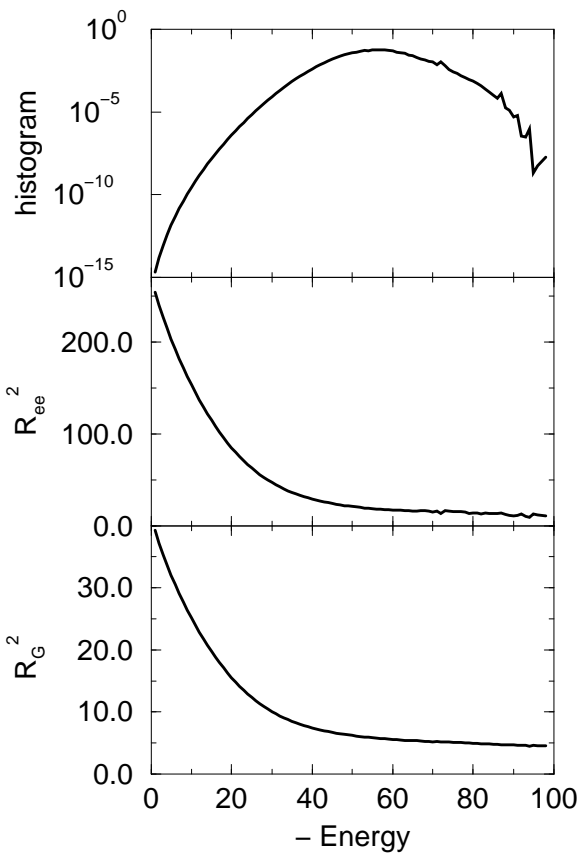, width=3truein}
\caption{ 
Histograms of (top) thermal weight, (middle) radius of gyration, $R^2_G$, 
and (bottom) end-to-end distance, $R^2_{ee}$, $vs$ energy $E$ 
for the 80-mer ``four helix bundle" at $T=0.75$.  }
\label{hist.vs.e}
\end{figure}

\begin{figure}
\begin{center}
\epsfig{figure=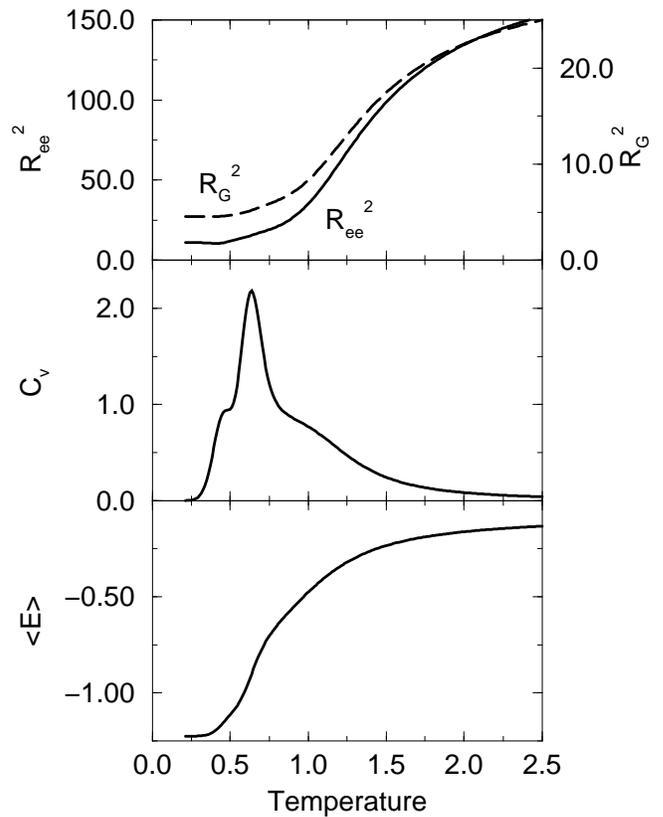, width=3.4truein}
\end{center}
\caption{
(top) Average end-to-end distance, $R^2_{ee}$, and radius of gyration, $R_G^2$,
(middle) specific heat per monomer, $C_v$, and average energy per monomer,
$<E>$,
 $vs$ temperature $T$ for the 80-mer ``four helix bundle".
}
\label{op.vs.T}
\end{figure}

\begin{figure}
\begin{center}
\epsfig{figure=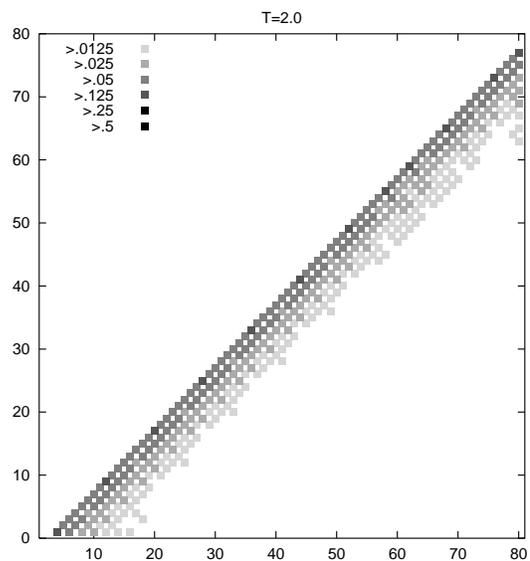, width=2.95truein}
\end{center}
\caption{ Thermally averaged contact matrix for the 80-mer ``four helix bundle" 
in the random coil phase ($T=2.0$). Different shades of grey denote different 
probabilities for the contact to exist.  }
\label{contact.200} 
\end{figure}

\noindent
of Ref. \cite{pekney}. While the new structure (Fig.~\ref{3d.struct.new})
is dominated by $\beta$ sheets, which can most clearly be seen in the 
contact matrix (see Fig.~\ref{3d.contact.new}), 
the structure in Fig.~\ref{3d.struct.old} 
is dominated by helices,
see also the corresponding contact matrix in Fig.~\ref{3d.contact.old}.

\begin{figure}
\begin{center}
\epsfig{figure=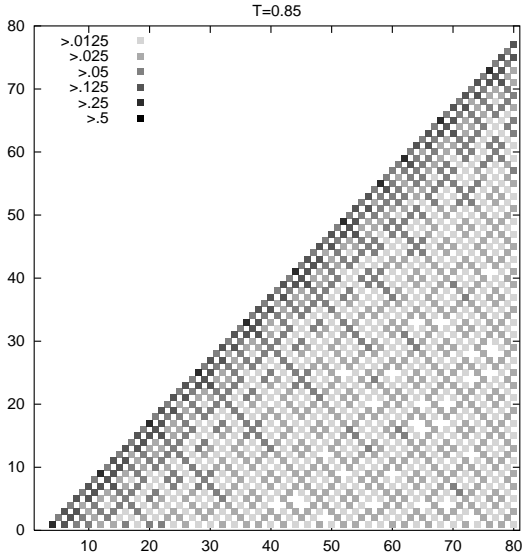, width=2.95truein}
\end{center}
\caption{ Thermally averaged contact matrix for the 80-mer ``four helix bundle" 
in the collapsed but unstructured phase ($T=0.85$).  }
\label{contact.85} 
\end{figure}

\begin{figure}
\begin{center}
\epsfig{figure=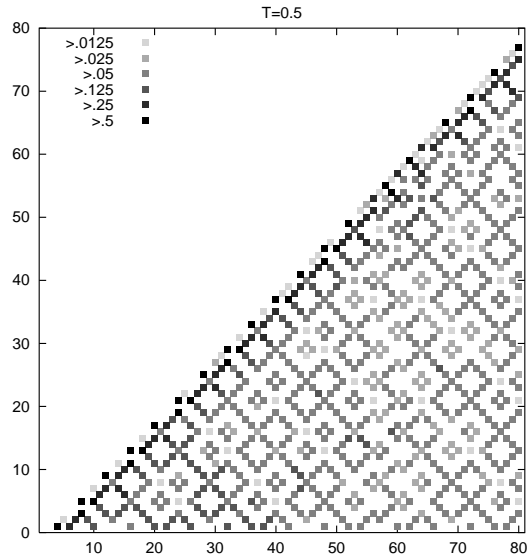, width=2.95truein}
\end{center}
\caption{ Thermally averaged contact matrix for the 80-mer ``four helix bundle" 
in the intermediate helix-dominated phase ($T=0.5$).
}
\label{contact.50} 
\end{figure}

\begin{figure}
\begin{center}
\epsfig{figure=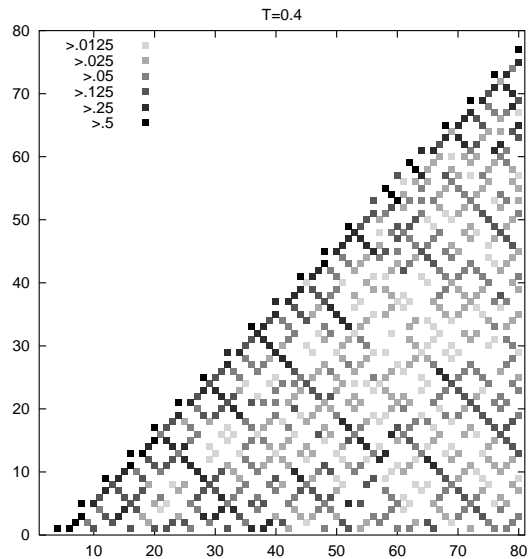, width=2.95truein}
\end{center}
\caption{ Thermally averaged contact matrix for the 80-mer ``four helix bundle" 
at the transition from the intermediate helix-dominated phase 
to the $\beta$-sheet dominated phase ($T=0.4$).
}  
\label{contact.40} 
\end{figure}

\begin{figure}
\epsfig{figure=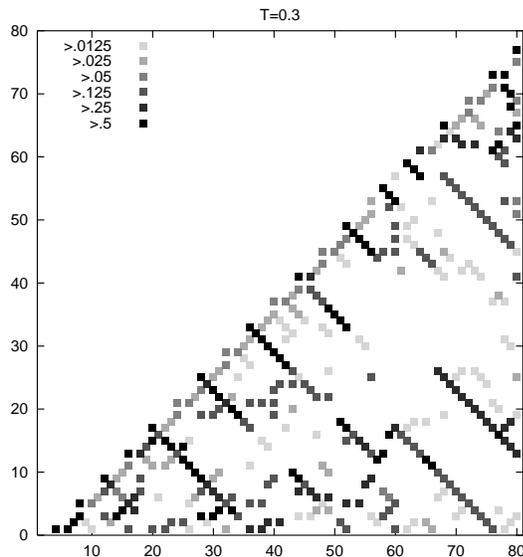, width=2.95truein}
\caption{ Thermally averaged contact matrix for the 80-mer ``four helix bundle" 
in the $\beta$-sheet dominated phase ($T=0.3$).
} 
\label{contact.30} 
\end{figure}

In order to analyze the folding transition of this sequence
we again constructed histograms of the distribution of energy,
end to end distance, and radius of gyration, by combining the results 
obtained at various temperatures between $T=0.45$ and 3. 
Figure~\ref{hist.vs.e} shows these distributions, reweighted so that it 
corresponds to $T=0.75$.
The thermal behavior of these order parameters as functions of $T$ is 
obtained by Laplace transform, and is shown in Fig.~\ref{op.vs.T}.
The behavior of energy, end-to-end distance and radius of gyration
follow closely each other and exhibit clearly only the smeared out
collapse of the chain from a random coil to some unstructured compact 
phase. In contrast, the specific heat exhibits more
structure: the shoulder around $T=1$ corresponds again to
the coil-globule collapse, but there are additional transitions seen 
around $T=0.62$ and $T=0.45$. The last one is the transition to 
the $\beta$-sheet dominated native phase. However, the transition 
at $T=0.62$ is from a unstructured globule 
to an intermediate phase that is helix-dominated but exhibits strong
tertiary fluctuations.
These structural transitions are illustrated in Figs~\ref{contact.200}
to \ref{contact.30} where the thermally averaged contact 
matrices are shown for the respective phases.

The intermediate, helix-dominated phase is particularly interesting.
To it apply some of the usual characteristics of a molten globule 
state \cite{Pain:93}: i) compactness, ii) large secondary structure content 
(although not necessarily native), and iii) strong fluctuations. This
qualifies it as a candidate for a molten globule state,
a phase that is absent in the folding transition of the HAP sequence 
\cite{prionnote}.

\subsection*{3d $\infty$ monomer types}
\label{sec:results:3dinfinity}

For all sequences with $N=27$ from \cite{Klimov:Thirumalai:96} 
we could reach the supposed ground state energies within $< 1$ hour. 
In no case we found energies lower than those quoted in 
\cite{Klimov:Thirumalai:96}, and we verified also the energies 
of low-lying excited states given in \cite{Klimov:Thirumalai:96}. 
Notice that these sequences were 
designed to be good folders by the authors of \hfill \cite{Klimov:Thirumalai:96}.

\noindent
This time the design had obviously been successful, which is mainly
due to the fact that the number of different monomer types is large. 
All sequences showed some 
gaps between the ground state and the bulk of low-lying states,
although these gaps are not very pronounced in some cases.

More conspicuous than these gaps was another feature: all low lying 
excited states were very similar to the ground state, as measured by
the fraction of contacts which existed also in the native configuration.
Stated differently, if the gaps were not immediately obvious, this was 
because they were filled by configurations which were very similar to 
the ground state and can therefore easily transform into the native state 
and back. Such states therefore cannot prevent a sequence from being 
a good folder. For none of the sequences of \cite{Klimov:Thirumalai:96} 
we found truely misfolded low-lying states with small overlap with the 
ground state.

\begin{figure}
\epsfig{figure=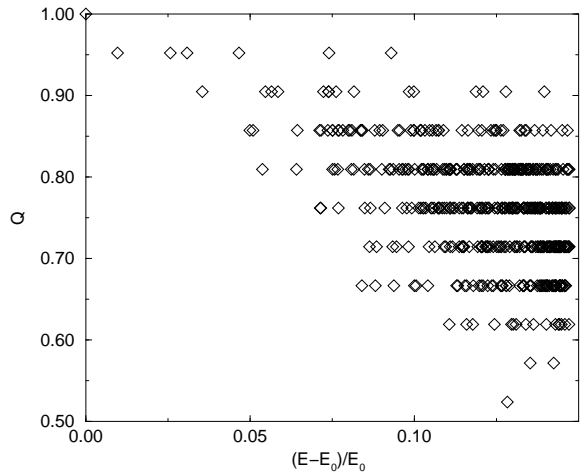, width=3truein}
\caption{ Overlap with ground state, $Q$, $vs$ energy $E$ of the lowest energy 
conformations
for sequence no. 70 of Ref. \protect\cite{Klimov:Thirumalai:96}.
}
\label{klimov.Q.vs.E.single} 
\end{figure}

\begin{figure}
\epsfig{figure=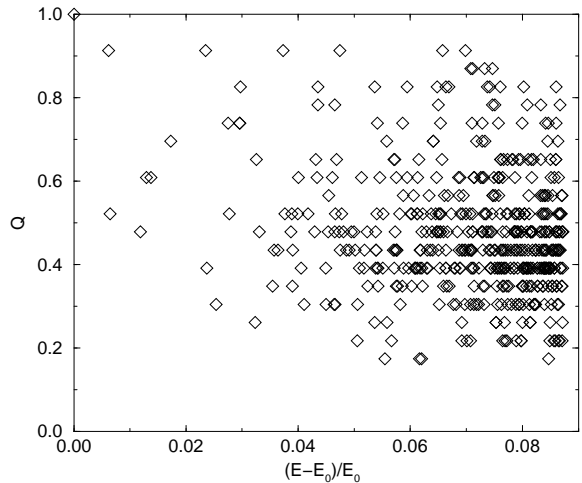, width=3truein}
\caption{ Overlap with ground state, $Q$, $vs$ energy $E$ of the lowest energy 
conformations for a single random sequence.
}
\label{klimov.Q.vs.E.random.single} 
\end{figure}

Figure~\ref{klimov.Q.vs.E.single} illustrates this feature for one particular 
sequence. There we show the {\it overlap} $Q$, defined as the fraction 
by non-bonded nearest-neighbor ground state contacts which exist also in 
the excited 
state, against the excitation energy. For this and for each of the following 
figures, the 500 lowest-lying states were determined. We see no low energy 
state with a small value of $Q$. To demonstrate that this is due to 
design, and is not a property of random sequences with the same potential 
distribution, we show in Fig.~\ref{klimov.Q.vs.E.random.single} the analogous 
distribution for a random sequence. 

\begin{figure}
\epsfig{figure=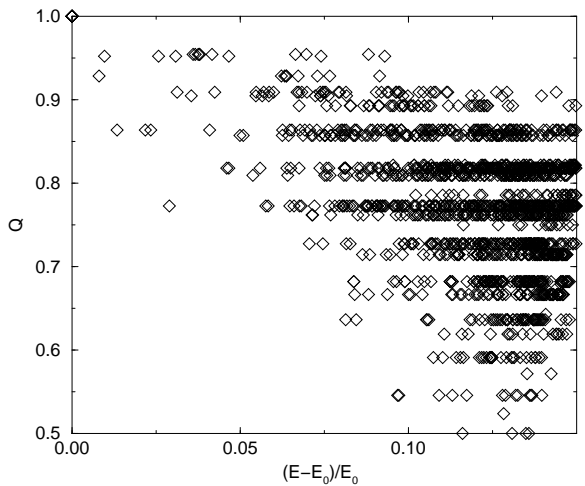, width=3truein}
\caption{ Overlap with ground state, $Q$, $vs$ energy $E$ of the lowest energy 
conformations
for sequences no. 61 to 70 of Ref. \protect\cite{Klimov:Thirumalai:96};
for better visibility, the same symbol is used for all sequences.
}
\label{klimov.Q.vs.E.total} 
\end{figure}

\begin{figure}
\epsfig{figure=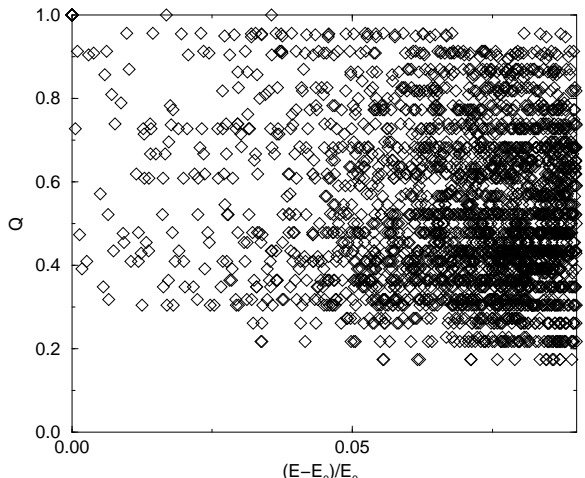, width=3truein}
\caption{ Overlap with ground state, $Q$, $vs$ energy of the lowest energy 
conformations
for ten random sequences;
for better visibility, the same symbol is used for all sequences.
}
\label{klimov.Q.vs.E.random.total} 
\end{figure}

To demonstrate that this difference is not merely due to a statistical 
fluctuation, we show in Fig.~\ref{klimov.Q.vs.E.total} 
the distributions for ten sequences from \cite{Klimov:Thirumalai:96} collected 
in a single plot. Since the ground state energies differ considerably for 
different sequences, we used 
normalized excitation energies $(E-E_0)/E_0$ on the x-axis. Analogous 
results for ten random sequences are shown in 
Fig.~\ref{klimov.Q.vs.E.random.total}.
While there is no obvious correlation between $Q$ and excitation energies 
for the random case, all low energy states with small $Q$ have been 
eliminated in the designed sequences (note the different ranges of $Q$ 
in Figs.~\ref{klimov.Q.vs.E.total} and \ref{klimov.Q.vs.E.random.total}). 

This elimination of truely misfolded low energy states without elimination of 
native-like low energy states might be an unphysical property of the 
design procedure used in \cite{Klimov:Thirumalai:96}, but we do not believe 
that this is the case. Rather,

\noindent
it should be a general feature of any design 
procedure, including the one due to biological evolution. It contradicts 
the claim of \cite{Sali:Shakhnovich:Karplus:94:Nature} that it is only the 
gap between native and first excited state which determines foldicity. On 
the other hand, our results are consistent with the ``funnel" scenario 
for the protein folding process \cite{Wolynes:Onuchic:Thirumalai:95},
where the folding pathway consists of states successively lower in energy 
and closer to the native state.

We note that for random sequences there are also excited 
states that have unit overlap with the native state, a feature not 
present in the folding sequences. These are cases where the native state 
has open loops and/or dangling ends, so that more compact conformations
have all contacts of the native state, but have -- in addition -- energetically 
unfavorable contacts resulting in a higher total energy.

\section*{Summary and Outlook}
\label{sec:summary}

We showed that the pruned-enriched Rosenbluth method (PERM) can be
very effectively applied to protein structure prediction in simple lattice
models. It is suited for calculating statistical properties and is very
successful in finding native states. In all cases it did better than 
any previous MC method, and in several cases it found lower energy states than 
those which had previously been conjectured to be native. 

We verified that ground states of the HP 
model are highly degenerate and have no gap, leading to bad folders. 
For sequences that are good folders we have established a funnel 
structure in state space: low-lying excited states of well-folding 
sequences have strong similarities to the ground state, while this is 
not true for non-folders with otherwise similar properties. 

Especially, we 
have presented a new candidate for the native 
configuration of a ``four helix bundle" sequence which had been studied 
before by several authors. 
The ground state structure of the ``four helix bundle" sequence,
being actually $beta$-sheet dominated,  
differs strongly from the helix-dominated intermediate phase.
This sequence, therefore, should not be a good folder. 

Although we have considered only lattice models in this paper, we 
should stress that this is not an inherent limitation of PERM. 
Straightforward extensions to off-lattice 
systems are possible and are efficient for homopolymers at 
relatively high temperatures \cite{alg}. Preliminary attempts to 
study off-lattice heteropolymers at low $T$ have not yet been particularly 
successful, but the inherent flexibility of PERM suggests several 
modifications which have not yet been investigated in detail. One 
of them are hybrid PERM-Metropolis approaches similar to that 
used for the HAP model in the present paper. Its success also suggests 
that similar hybrid approaches should be useful for models with 
more complicated monomers. Another improvement of PERM which could 
be particularly useful for off-lattice simulations might consist in 
more sophisticated algorithms for positioning the monomers when assembling 
the chain.  Work along these lines is in progress, and we hope to 
report on it soon.

\acknowledgements

The authors are grateful to Gerard Barkema for helpful discussions 
during this work. One of them (P.G.) wants to thank also Eytan Domany 
and Michele Vendruscolo for very informative discussions, and to 
Drs. D.K. Klimov and R. Ramakrishnan for correspondence.

\end{multicols}


\begin{references}
  
\bibitem{Gierasch:King:90} L. M. Gierasch and J. King (eds.), 
{\it Protein Folding, Deciphering the Second Half of the Genetic Code} 
(AAAS, New York, 1990)

\bibitem{Creighton:92} T.E. Creighton (ed.), {\it Protein Folding} (Freeman, 
  New York, 1992)

\bibitem{Merz:LeGrand:94} 
K. M. Merz Jr. and S. M. LeGrand (eds.), 
{\it The Protein Folding Problem and Tertiary Structure Prediction} 
(Birkh\"auser, Boston, 1994)

\bibitem{Brunak:95} 
H. Bohr and S. Brunak(eds.), {\it Protein Folds: A Distance Based Approach}
(CRC Press, Boca-Raton/FL. 1996).

\bibitem{Drexler:86} 
{K.~E.~Drexler},~{\it Engines~of~Creation}, (Anchor Books, 1986); 
available also on the web at the URL
http://www.asiapac.com/EnginesOfCreation/.

\bibitem{Gross:95} 
M. Gro\ss, {\it Expeditionen in den Nanokosmos}, (Birkh\"auser, Basel, 1995).

\bibitem{Crippen:Maiorov:94} 
G. M. Crippen and V. N. Maiorov, in \protect\cite{Merz:LeGrand:94}, p.231-277.

\bibitem{Kolinski:Skolnick:95} 
A. Kolinski and J. Skolnick, {\it Lattice Models of Protein Folding,
Dynamics and Thermodynamics},
(Chapman \& Hall, New York, 1996).

\bibitem{Mirny:Shakhnovich:96}
L. A. Mirny and E. I. Shakhnovich, J. Mol. Biol. {\bf 264} 1164 (1996).


\bibitem{Sali:Shakhnovich:Karlplus:94:JMB}
A. Sali, E. I. Shakhnovich and M. Karplus J. Mol. Biol. {\bf 235} 1614 (1994).

\bibitem{Hansmann:Okamoto:94-96}
U. H. E. Hansmann and Y. Okamoto, J. Comp. Chem. {\bf 14}, 1333 (1993);
Physica A {\bf 212}, 415 (1994); Phys. Rev. E {\bf 54}, 5863 (1996).

\bibitem{Wilson:Cui:94}
S. R. Wilson and W. Cui, in \protect\cite{Merz:LeGrand:94}, p.43-70.

\bibitem{Unger:Moult:93} R. Unger and J. Moult, J. Mol. Biol. {\bf 231}, 75 
(1993)

\bibitem{LeGrand:Merz:94a}
S. M. LeGrand and K. M. Merz jr., in \protect\cite{Merz:LeGrand:94}, p.109-124.

\bibitem{Dill:Fiebig:Chan:93}
K. A. Dill, K. M. Fiebig and H. S. Chan, Proc. Natl. Acad. Sci USA {\bf 90}, 
1942 (1993).

\bibitem{Eisenhaber:Persson:Argos:95}
F. Eisenhaber, B. Persson and P. Argos,
Crit. Rev. Biochem. Mol. Biol. {\bf 30}, 1 (1995).

\bibitem{alg} P.~Grassberger, Phys. Rev. {\bf E}, in press (1997)
  
\bibitem{rr} M.N.~Rosenbluth and A.W.~Rosenbluth, J.~Chem.~Phys. {\bf 23}, 256 
(1955)
  
\bibitem{Frauenkron:etal:97} 
H. Frauenkron, U. Bastolla, E. Gerstner, P. Grassberger 
and W. Nadler, submitted (1997).

\bibitem{kantor-kardar:94} Y. Kantor and M. Kardar, Europhys. Lett. 
   {\bf 28}, 169 (1994)  

\bibitem{Dill:85} K.A.~Dill, Biochemistry {\bf 24}, 1501 (1985) 

\bibitem{Dill:89-91} K.F.~Lau and K.A.~Dill, Macromolecules {\bf 22}, 
  3986 (1989); J. Chem. Phys. {\bf 95}, 3775 (1991); 
  H.S.~Chan, and K.A.~Dill, J. Chem. Phys. {\bf 95}, 3775 (1991)

\bibitem{socci1} N.D.~Socci and J.N. Onuchic, J.~Chem.~Phys. {\bf 101},
  1519 (1994)

\bibitem{otoole} E.~O'Toole and A.~Panagiotopoulos, J.~Chem.~Phys.
  {\bf 97}, 8644 (1992)

\bibitem{Krausche:Nadler:97a} 
T. Krausche and W. Nadler, to be published (1997).

\bibitem{Sali:Shakhnovich:Karplus:94:Nature}
A. Sali, E. I. Shakhnovich and M. Karplus, Nature {\bf 369} 248 (1994).

\bibitem{Klimov:Thirumalai:96} D.K. Klimov and D. Thirumalai, PROTEINS: 
  Structure, Function and Genetics {\bf 26}, 411 (1996); sequences are 
  available from http://www.glue.umd.edu/$\sim$klimov.

\bibitem{pekney} R.~Ramakrishnan, B.~Ramachandran and J.F.~Pekney,
  J.~Chem.~Phys. {\bf 106}, 2418 (1997)

\bibitem{deutsch} J.M.~Deutsch, J. Chem. Phys. {\bf106}, 8849 (1997).
  
\bibitem{yue-shak} K.~Yue {\it et al.}, Proc.~Natl.~Acad.~Sci.~USA
  {\bf 92}, 325 (1995)

\bibitem{Shakhnovich:93-94}
E. I. Shakhnovich and A. M. Gutin, Proc. Natl. Acad. Sci USA {\bf 90}, 7195 
(1993);
E. I. Shakhnovich, Phys. Rev. Lett. {\bf 72}, 3907 (1994).

\bibitem{Yue:Dill:95}
K. Yue and K. A. Dill, Proc. Natl. Acad. Sci USA {\bf 92}, 146 (1995).

\bibitem{Ebeling:Nadler:95-97} M. Ebeling and W. Nadler,
Proc. Natl. Acad. Sci. USA {\bf 92}, 8798 (1995);
Biopolymers {\bf 41}, 165 (1997).

\bibitem{deutsch-kurosky:96} J.M. Deutsch and T. Kurosky, 
Phys. Rev. Lett. {\bf 76}, 323 (1996).

\bibitem{kremer} J. Batoulis and K. Kremer, J. Phys. {\bf A 21}, 127 (1988) 

\bibitem{garel} T. Garel and H. Orland, J. Phys. {\bf A 23}, L621 (1990)

\bibitem{velicson} B. Velikson, T. Garel, J.-C. Niel, H. Orland and 
     J.C. Smith, J. Comput. Chem. {\bf 13}, 1216 (1992)

\bibitem{umrigar} C.J. Umrigar, M.P. Nightingale, and K.J. Runge, 
  J. Chem. Phys. {\bf 99}, 2865 (1993) 

\bibitem{enrich} F.T. Wall and J.J. Erpenbeck, J. Chem. Phys. {\bf 30},
  634, 637 (1959)

\bibitem{multic} H. Frauenkron and P. Grassberger, preprint
   cond-mat/9707101 (1997)

\bibitem{stiff} U. Bastolla and P.Grassberger, preprint 
   cond-mat/9705178 (1997)

\bibitem{spiral} G. T. Barkema, U. Bastolla, and P. Grassberger, preprint 
   cond-mat/9707312 (1997)

\bibitem{Matheson:Scheraga:78}
R. R. Matheson and H. A. Scheraga, Macromolecules {\bf 11}, 819 (1978).

\bibitem{Grassberger:unpublished}
At first sight, one might believe that allowing the chain to grow
at both ends should decrease the attrition rate and hence be 
advantageous. For homopolymers one can see easily that this is not 
true. Attrition is actually decreased, since chains which have one 
end blocked can still grow at the other end. But such chains have 
small Rosenbluth factors and thus extremely low $W$ in average. 
Therefore, they just cost efforts without 
efficiently contributing to statistical averages. 
For heteropolymers this argument no longer holds, however, and
it is not 
clear why growing the chain at both ends is not efficient either. 




\bibitem{enumerationNote}
Note, however, that there exists an alternative approach:
for a fixed chain geometry 
a computationally effective ( $O(N)$ instead of $O(2^N)$ )
complete enumeration of all side group conformations is possible;
E. Gerstner, P. Grassberger and W. Nadler, to be published.

\bibitem{Finkelstein:Gutin:Badretdinov:95}
A. V. Finkelstein, A. M. Gutin and A. Y. Badretdinov
PROTEINS: Structure, Function and Genetics {\bf 23}, 151 (1995).

\bibitem{Shakhnovich:Gutin:90-93}
E. I. Shakhnovich and A. M. Gutin, J. Chem. Phys. {\bf93} 5967 (1990);
A. M. Gutin and E. I. Shakhnovich, J. Chem. Phys. {\bf98} 8174 (1993).

\bibitem{mchisto} A.M.~Ferrenberg and R.H.~Swendsen, Phys.~Rev.~Lett.
  {\bf 61}, 2635 (1988); {\bf 63}, 1195 (1989)

\bibitem{Bennet:76}
C. H. Bennet, J. Comp. Phys. {\bf 22}, 245 (1976).

\bibitem{Chan:Dill:89-90}
H. S. Chan and K. A. Dill,
Macromolecules {\bf 22} 4559 (1989);
J. Chem. Phys. {\bf 92}, 3118 (1990).

\bibitem{Pain:93}
H. Christensen and R. H.Pain,
Europ. Biophys. J. {\bf 19}, 221 (1991).

\bibitem{prionnote}
In \cite{deutsch}, indications for the $\beta$-sheet phase were found, too, 
and on this basis an analogy to the competition of structures in the prion 
problem [see, e.g., J. Nguyen {\it et al.}, Biochemistry {\bf 34}, 4186 (1995)] 
was drawn. However, our results showing that the helical phase is strongly 
structurally disordered demonstrate that this analogy cannot be perfect. In 
a model for prions, both competing phases should be much more well defined.

\bibitem{Wolynes:Onuchic:Thirumalai:95}
P. G. Wolynes, J. N. Onuchic, and D. Thirumalai,
Science {\bf 267}, 1619 1995).


\end{references}
\end{document}